\documentclass[twocolumn]{aastex631}
\usepackage{color}
\usepackage{multirow}
\usepackage{amsmath}
\usepackage{booktabs}

\newcommand{\JKCS}{JKCS~041}
\newcommand{\XLSS}{XLSSC~122}

\newcommand{\LCDM}{$\Lambda$CDM}
\newcommand{\HST}{HST}
\newcommand{\mytilde}{\raise.19ex\hbox{$\scriptstyle\sim$}}

\newcommand{\solarm}{$10^{14}~M_{\sun}$}
\newcommand{\persqarcmin}{arcmin$^{-2}$}
\newcommand{\sqarcmin}{arcmin$^{2}$}

\shorttitle{WL Study of JKCS~041 and XLSSC~122}
\shortauthors{Kim et al.}

\begin{document}

\title{Exploring the Masses of the Two Most Distant Gravitational Lensing Clusters at the Cosmic Noon}

\author{Jinhyub Kim}
\affiliation{Department of Physics, University of Oxford, Denys Wilkinson Building, Keble Road, Oxford OX1 3RH, UK; \texttt{jinhyub.kim@physics.ox.ac.uk}}

\author{M. James Jee}
\affiliation{Department of Astronomy, Yonsei University, 50 Yonsei-ro, Seoul 03722, Korea; \texttt{mkjee@yonsei.ac.kr}}
\affiliation{Department of Physics, University of California, Davis, One Shields Avenue, Davis, CA 95616, USA}

\author{Stefano Andreon}
\affiliation{INAF–Osservatorio Astronomico di Brera, via Brera 28, 20121, Milano, Italy}

\author{Tony Mroczkowski}
\affiliation{European Southern Observatory (ESO), Karl-Schwarzschild-Strasse 2, Garching 85748, Germany}

\author{Lance Miller}
\affiliation{Department of Physics, University of Oxford, Denys Wilkinson Building, Keble Road, Oxford OX1 3RH, UK; \texttt{jinhyub.kim@physics.ox.ac.uk}}

\author{Joshiwa van Marrewijk}
\affiliation{European Southern Observatory (ESO), Karl-Schwarzschild-Strasse 2, Garching 85748, Germany}
\affiliation{Leiden Observatory, Leiden University, P.O. Box 9513, 2300 RA Leiden, The Netherlands}

\author{Hye Gyeong Khim}
\affiliation{Department of Astronomy, Yonsei University, 50 Yonsei-ro, Seoul 03722, Korea; \texttt{mkjee@yonsei.ac.kr}}

\begin{abstract}

Observations over the past decade have shown that galaxy clusters undergo the most transformative changes during the $z = 1.5 - 2$ epoch. However, challenges such as low lensing efficiency, high shape measurement uncertainty, and a scarcity of background galaxies have prevented us from characterizing their masses with weak gravitational lensing (WL) beyond the redshift $z\sim1.75$. 
In this paper, we report the successful WL detection of JKCS 041 and XLSSC 122 at $z=1.80$ and $z=1.98$, respectively, utilizing deep infrared imaging data from the Hubble Space Telescope with careful removal of instrumental effects. These are the most distant clusters ever measured through WL.
The mass peaks of JKCS~041 and XLSSC~122, which coincide with the X-ray peak positions of the respective clusters, are detected at the $\mytilde3.7\sigma$ and $\mytilde3.2\sigma$~levels, respectively. Assuming a single spherical Navarro-Frenk-White profile, we estimate that JKCS~041 has a virial mass of $M_{200c} = (5.4\pm1.6)~\times~10^{14}~M_{\sun}$ while the mass of XLSSC~122 is determined to be $M_{200c} = (3.3\pm1.8)~\times~10^{14}~M_{\sun}$. These WL masses are consistent with the estimates inferred from their X-ray observations. We conclude that although the probability of finding such massive clusters at their redshifts is certainly low, their masses can still be accommodated within the current $\Lambda$CDM paradigm.

\end{abstract}




\section{Introduction} \label{chap1}
As the largest gravitationally bound objects detached from the Hubble flow, galaxy clusters form at the intersections of cosmic filaments, growing via the gravitational accretion of matter over cosmic time. They serve as powerful astrophysical laboratories where insights into plasma physics, galaxy evolution, and dark matter properties can be gained. In addition, their aggregate statistics are invaluable diagnostics of the growth rate of the large-scale structure of the universe.

Cluster number densities per mass bin, referred to as the mass function, are sensitive to cosmological parameters, in particular, the matter density $\Omega_M$ and the normalization of the matter power spectrum $\sigma_8$ (e.g., \citealt{Borgani2011}; \citealt{Planelles2015}). However, since the effects of the two parameters $\Omega_M$ and $\sigma_8$ are highly degenerate, it is necessary to use a wide redshift baseline to break the degeneracy (e.g., \citealt{Bahcall1997}; \citealt{Bahcall1998}; \citealt{Sartoris2016}). 
Consequently, increasing efforts have been made to find and measure galaxy clusters at higher and higher redshifts, driven by a range of scientific objectives, including constraining cosmological parameters and understanding cluster formation and evolution.

A number of campaigns have been conducted to find high-$z$ clusters and to determine their masses through galaxy, X-ray, and Sunyaev-Zel’dovich (SZ) observations (e.g., \citealt{Fassbender2011}; \citealt{Mehrtens2012}; \citealt{Bleem2015}; \citealt{Hilton2021}).
These surveys are efficient at detecting high-$z$ clusters. 
However, mass estimation from these observables is prone to bias, as it assumes hydrostatic equilibrium or scaling relations that can vary with cluster redshift (e.g., \citealt{Nelson2012}; \citealt{Rasia2014}).

Weak gravitational lensing (WL) is a powerful tool to measure accurate cluster masses since the lensing signal does not depend on the dynamical state of the lens. Thus, many cosmological studies based on cluster mass function calibrate the mass using WL results. However, WL studies of high-redshift clusters are difficult in several aspects. 
First, high-redshift lenses ($z_{\rm lens} > 1$) are less effective than intermediate-redshift lenses. For instance, when the source is at $z_{\rm source}\sim2$, the lensing efficiency peaks at $z_{\rm lens}\sim0.5$. 
Beyond this point, the efficiency declines, making the lensing-induced distortion weaker at higher lens redshifts for a given lens mass. Second, the lensing signal comes from very distant, faint, and small galaxies, whose shapes are hard to determine precisely. Third, the source density decreases substantially because only the sources behind the already high-redshift cluster are lensed.

Space-based observations can overcome these challenges by providing higher-resolution images with smaller point spread functions (PSFs) than ground-based observations, enabling us to measure subtle distortions in much fainter and smaller background galaxy images more reliably. In addition, significant advantages are gained when using infrared imaging data for high-redshift WL, as source galaxy shapes at high redshifts ($z>1$) are easier to characterize in IR than in optical; the surface brightness is higher, and the light profile is smoother. Indeed, most existing WL studies of high-$z$ galaxy clusters at $z > 1$ relied on Hubble Space Telescope (\HST) imaging data (e.g., \citealt{Jee2011}; \citealt{Schrabback2018}; \citealt{Kim2019}), and a substantial fraction of lenses at $z>1.4$ have been measured with IR imaging data. Nevertheless, the number of clusters measured with WL rapidly drops with redshift. For instance, at $z>1.7$, only three galaxy clusters have been probed by WL to date: SPT-CL J0459-4947 at $z=1.71$ (SPT0459; \citealt{Zohren2022}), SpARCS1049+56 at $z=1.71$ (SpARCS1049; \citealt{Finner2020}), and IDCS J1426+3508 at $z=1.75$ (IDCS1426; \citealt{Jee2017}), with the latter IDCS1426+3508 at $z=1.75$ currently being the highest redshift lensing cluster on record.

Studies of high-$z$ galaxy clusters at $z\gtrsim1.5$ over the past decade presented many remarkable findings on the cluster galaxies and their environments. While the $1.5\lesssim z \lesssim 2$ epoch may be the most transformative (i.e., transition from proto-cluster to cluster) phase (e.g., \citealt{Overzier2016}; \citealt{Wang2016}; \citealt{Chiang2017}) in cluster evolution for the average population, many observations (e.g., \citealt{Andreon2014}; \citealt{Newman2014}; \citealt{Noordeh2021}; \citealt{Mei2023}) indicate that in some high-redshift clusters, mature (early-type) galaxies dominate the cluster population, as they do at the present epoch. 
Additionally, contrary to the theoretical prediction, abundant intracluster light has been detected in this epoch (\citealt{Joo2023}; \citealt{Werner2023}), with a fraction comparable to that found in their low-$z$ counterparts. Also, X-ray observations show that some clusters in this epoch possess X-ray emitting intracluster medium, which is typically used as a criterion to distinguish proto-clusters and clusters (e.g., \citealt{Overzier2016}). However, conspicuously missing is our understanding of their dark matter potential well, which is one of the most critical environmental factors.

We present WL masses of \JKCS~and \XLSS~at $z=1.80$ and $z=1.98$, respectively, based on HST Wide Field Camera 3 Infrared (WFC3/IR) imaging. They represent the most distant clusters measured with WL. \JKCS~was discovered in 2006 from the J and K observations of the UKIRT Infrared Deep Sky Survey \citep{Lawrence2007} Early Data Release \citep{Dye2006} as a clustering of sources of similar colors \citep{Andreon2009}. \cite{Andreon2009} also revealed a diffuse X-ray emission with a temperature of $7.4_{-3.3}^{+5.3}$~keV from the Chandra observation. The MUSTANG-2 SZ observation found that the cluster has a low central pressure with the SZ peak $\mytilde26$\arcsec~offset from the X-ray peak \citep{Andreon2023}. 
On the other hand, \XLSS~was originally detected as an extended X-ray source in the XMM Large Scale Structure survey \citep{Pierre2004} and was later found to coincide with a compact over-density of galaxies with $z_{\rm phot}=1.9 \pm 0.2$ \citep{Willis2013}. From the \HST~G141 grism data, \cite{Willis2020} confirmed 37 member galaxies at a mean redshift of 1.98 and demonstrated that \XLSS~is indeed a surprisingly mature cluster with evolved members when the universe was less than a quarter of its current age. 
Multiwavelength studies indicate that both \JKCS~and \XLSS~may be massive clusters ($ \gtrsim 1.5 \times10^{14}~M_{\sun}$, \citealt{Andreon2014}; \citealt{vanMarrewijk2024}) for their redshifts. However, their masses have yet to be confirmed through gravitational lensing, as its signal does not depend on their unknown dynamical states.

This paper is organized as follows. In \S\ref{chap2}, we describe our \HST~WFC3/IR imaging data, their reduction, and the WL analysis procedure. We present our mass reconstruction and estimation in \S\ref{section_results}. In \S\ref{section_discussion}, the WL masses are compared with the results from previous non-lensing studies before the summary is presented in \S\ref{section_summary}.

Throughout the paper, we assume a flat $\Lambda$CDM cosmology with $\Omega_M=1-\Omega_{\Lambda}=0.3$ and $h=0.7$ to interpret the WL signal. We use the AB magnitude system corrected for the Milky Way foreground extinction, and quoted uncertainties are at the $1\sigma$ ($\mytilde68.3$\%) level. We represent the cluster mass by $M_{\Delta_c}$, the mass enclosed within a spherical radius inside which the mean density equals $\Delta_c$ times the critical density of the universe at the cluster redshift.

\begin{figure*}
\centering
\includegraphics[width=170mm]{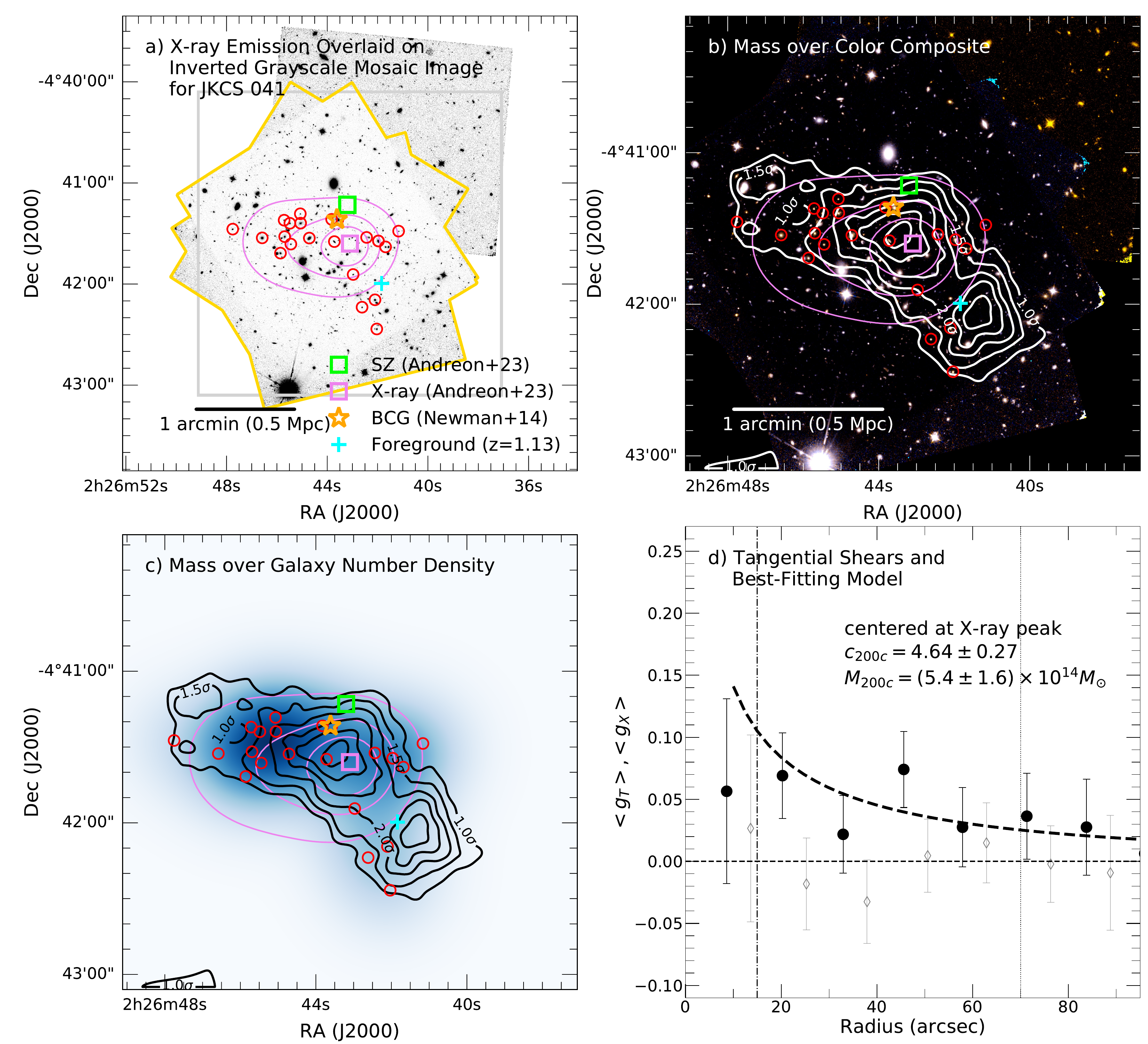}
\caption{WL results of \JKCS. Our two-dimensional (2D) mass reconstruction is obtained by {\tt FIATMAP}. We denote the SZ peak with a green square, the X-ray peak with a violet square, the brightest cluster galaxy (BCG) position with an orange star, the foreground structure center at $z=1.13$ with a cyan cross, and the spectroscopic members with red circles.
\textbf{(a)} The footprint of the \HST~observation overlaid with X-ray emission contours (violet; \citealt{Andreon2023}). The inverted grayscale represents the intensity from the WFC3/IR F160W filter, while the yellow box outlines the footprint of the WFC3/IR F105W observation. The mass reconstruction is limited to the central $\sim1.5\times1.5$ arcmin$^2$ region (gray box).
\textbf{(b)} Mass (white) and X-ray emission (violet) contours overlaid on the color composite. This color composite is created using \HST~WFC3 IR/F105W (blue), the mean of WFC3 IR/F105W and WFC3 IR/F160W (green), and WFC3 IR/F160W (red). The outermost mass contour corresponds to $1\sigma$ significance, with significance increasing inward by $0.5\sigma$, reaching a peak significance value of $\mytilde3.7\sigma$. 
\textbf{Our WL mass centroid is consistent with the X-ray peak.} 
\textbf{(c)} Mass (black) and X-ray emission (violet) contours overlaid on the galaxy number density map of the confirmed cluster member galaxies from \citet{Newman2014}. The smoothing scale of the number density map is FWHM~$\sim33\arcsec$. Overall, correlations are good between the galaxy number density map, the X-ray contours, and our WL mass contours.
\textbf{(d)} Reduced tangential shear profile (filled circles) with respect to the X-ray peak. The cross shear profile (open diamonds), obtained by rotating the source galaxy images by 45\degr, is consistent with zero. The best-fit model under a single spherical NFW halo assumption with the mass-concentration relation of \cite{DJ19} is denoted with dashed line. The dot-dashed vertical line at $r\sim15\arcsec$ indicates the cutoff radius inside which the signal is ignored in our NFW profile fitting. The dotted vertical line at $r\sim70\arcsec$ marks the maximum radius where the azimuthal average can be obtained from a complete circle.}
\label{fig:mass_map_JKCS041}
\end{figure*}

\begin{figure*}
\centering
\includegraphics[width=170mm]{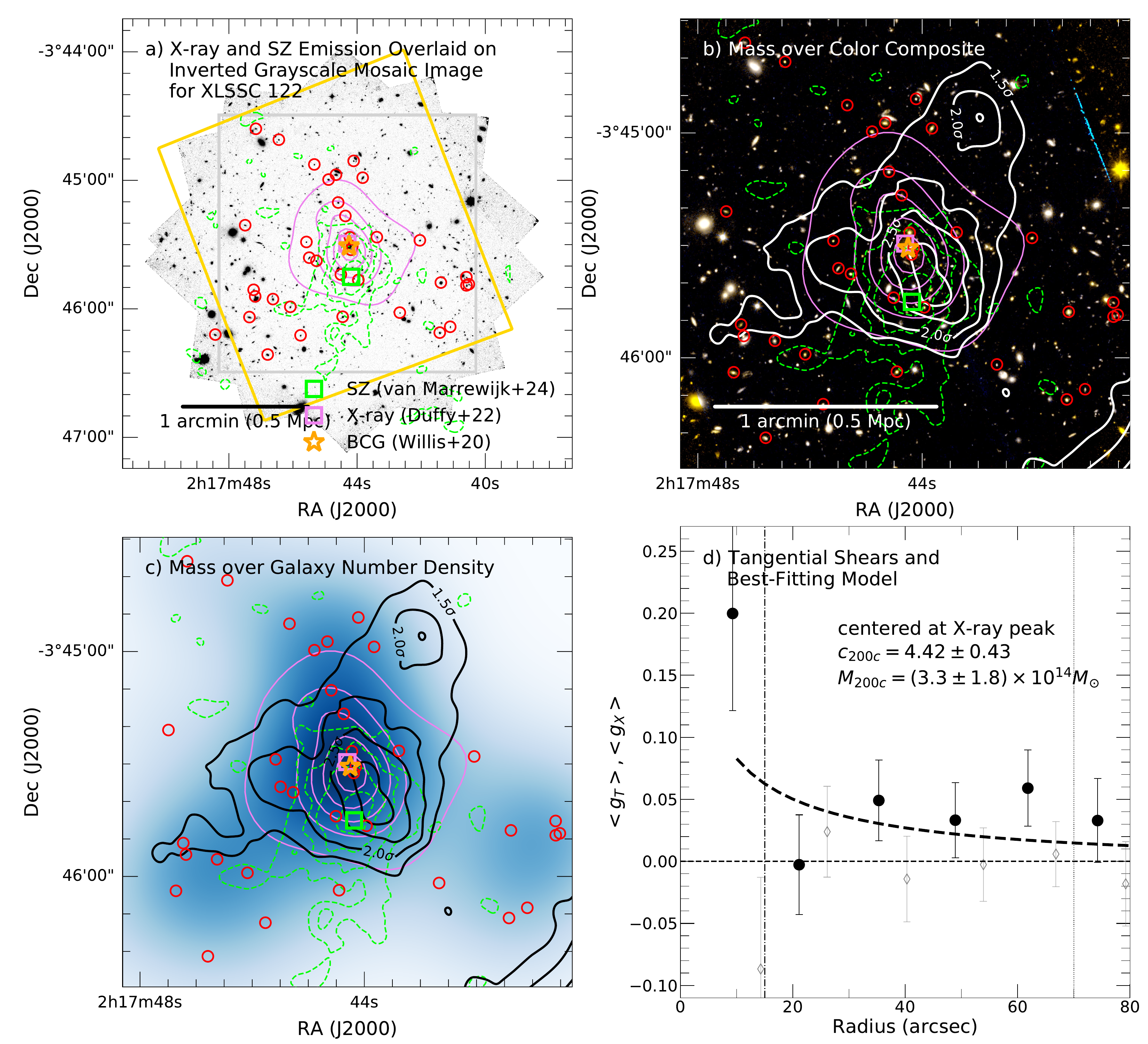}
\caption{Same as Figure~\ref{fig:mass_map_JKCS041} but for \XLSS.
\textbf{(a)} The footprint of the \HST~observation overlaid with X-ray (violet; \citealt{vanMarrewijk2024}) and SZ (green; \citealt{vanMarrewijk2024}) contours. The SZ peak, X-ray peak, and BCG position are indicated with a green square, a violet square, and an orange star, respectively. The inverted grayscale represents the intensity in F140W. The yellow box outlines the footprint of the F105W observation. Our weak lensing mass reconstruction is performed on the central $\mytilde1\times1$ arcmin$^2$ region marked by a gray box.
\textbf{(b)} Mass (white), X-ray emission (violet), and SZ (green) contours overlaid on the color composite, which is created with F105W (blue), F105W+F140W (green), and F140W (red). The outermost mass contour corresponds to $1.5\sigma$ significance, with significance increasing inward by $0.5\sigma$, reaching a peak significance value of $\mytilde3.2\sigma$. Our WL mass centroid is consistent with the SZ, X-ray, and BCG centers. 
\textbf{(c)} Mass (black), X-ray emission (violet), and SZ (green) contours overlaid on a galaxy number density map created with the confirmed cluster member galaxies from \citet{Willis2020}. 
\textbf{(d)} Reduced tangential shear profile with respect to the X-ray peak. The best-fit single spherical NFW model is indicated with the dashed line.}
\label{fig:mass_map_XLSSC122}
\end{figure*}

\section{\HST~Observations and Weak-Lensing Analysis} \label{chap2}
\subsection{\HST~Observations} \label{section_observations}
We employ three \HST~programs in our WL analysis: PROP 12927 (PI: A. Newman) and PROP 12990 (PI: A. Muzzin) for \JKCS, and PROP 15267 (PI: R. Canning) for \XLSS. 
PROP 12927 provides grism spectroscopy and imaging data with two WFC3/IR filters, F105W and F160W. The WFC3/IR images from this program are centered on the X-ray peak of \JKCS~(Figure~\ref{fig:mass_map_JKCS041}). The total integration times are 2671 s and 4509 s for F105W and F160W, respectively. PROP 12990 targeted galaxies $\mytilde1\arcmin$ northwest of \JKCS~and took a shallow (integration time of 812 s) and single-band (F160W) image. The total field size of the F160W observation is $\mytilde8.8$~\sqarcmin, while the F105W image covers $\mytilde6.3$~\sqarcmin. 
PROP 15267 includes grism spectroscopy and WFC3/IR imaging with F105W and F140W. The F105W image is acquired in a single visit, with a total integration of 2612~s, covering an area of $\mytilde4.7$~\sqarcmin~ (Figure~\ref{fig:mass_map_XLSSC122}). The F140W image is slightly wider ($\mytilde6.2$~\sqarcmin) and deeper (total integration of 5170~s).

Because of the difference in the coverage, there are regions without color information. Hereafter, we refer to the region where color information is available as ``Region A" (a common area enclosed by the yellow line in the upper left panels of Figures~\ref{fig:mass_map_JKCS041} and~\ref{fig:mass_map_XLSSC122}). The area outside ``Region A” is referred to as ``Region B” where only the F140W filter (for \XLSS) or the F160W filter (for \JKCS) is available.

We reduced the \HST~WFC3/IR data following the procedures outlined in \cite{Jee2017} and \cite{Kim2019, Kim2021}. Briefly, we first aligned the individual {\tt FLT} exposures with respect to a reference frame by obtaining astrometric solutions based on common astronomical sources in the overlapping regions. Next, we used the MultiDrizzle software \citep{2002multidrizzle} to remove cosmic rays and sky background and to combine all exposures. We chose the Gaussian kernel with a {\tt pixfrac} parameter of $0.7$ and an output pixel scale of 0\farcs05~per pixel. As a sanity check, we repeated the data reduction using {\tt TweakReg} and {\tt AstroDrizzle} in the DrizzlePac package \citep{DrizzlePac} and confirmed that the outputs are virtually identical.

Following the data reduction process, we detected objects with SExtractor \citep{Bertin1996} in dual-image mode. We created the detection image by stacking all exposures according to their weights, and we identified objects if they have at least ten connected pixels above 1.5 times the sky rms.

\begin{figure*}
\centering
\includegraphics[width=8.5cm]{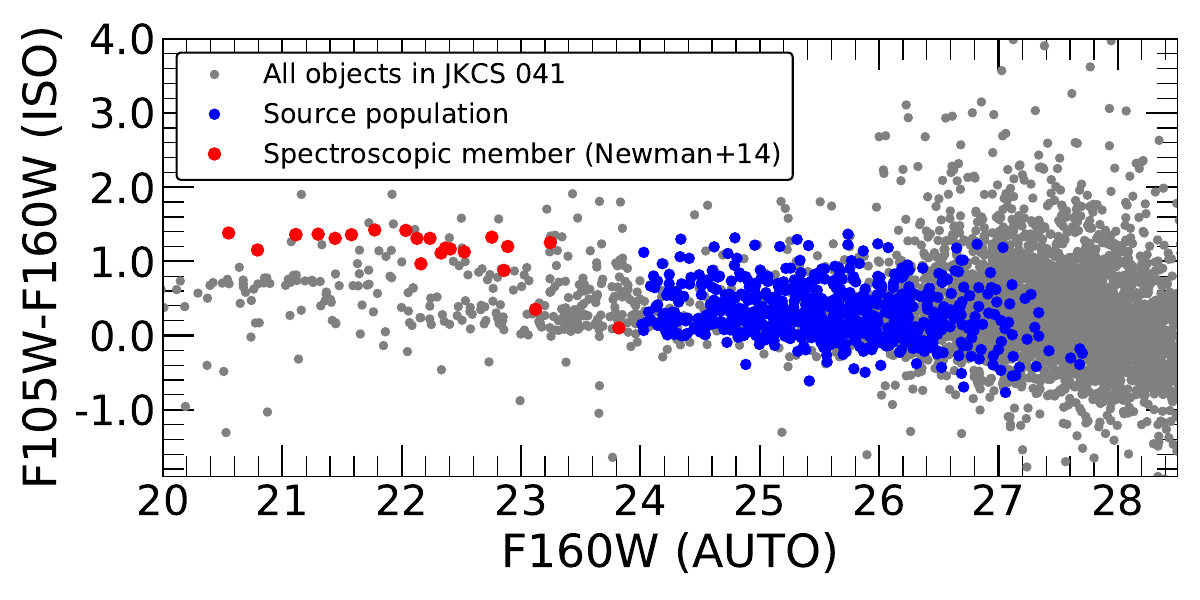}
\includegraphics[width=8.5cm]{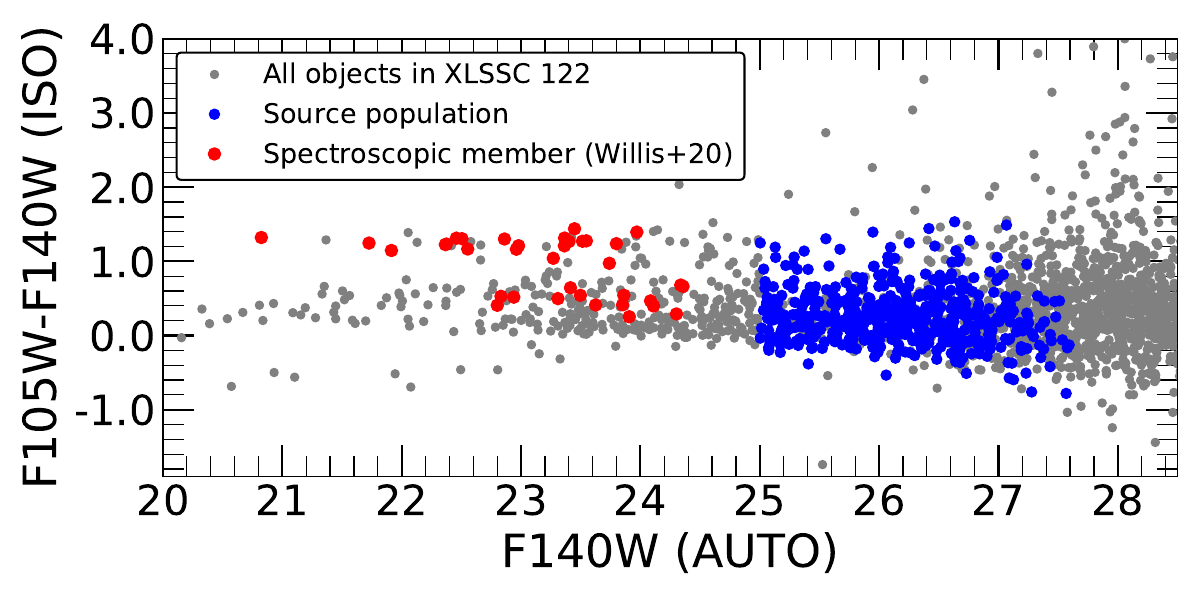}
\includegraphics[width=8.5cm]{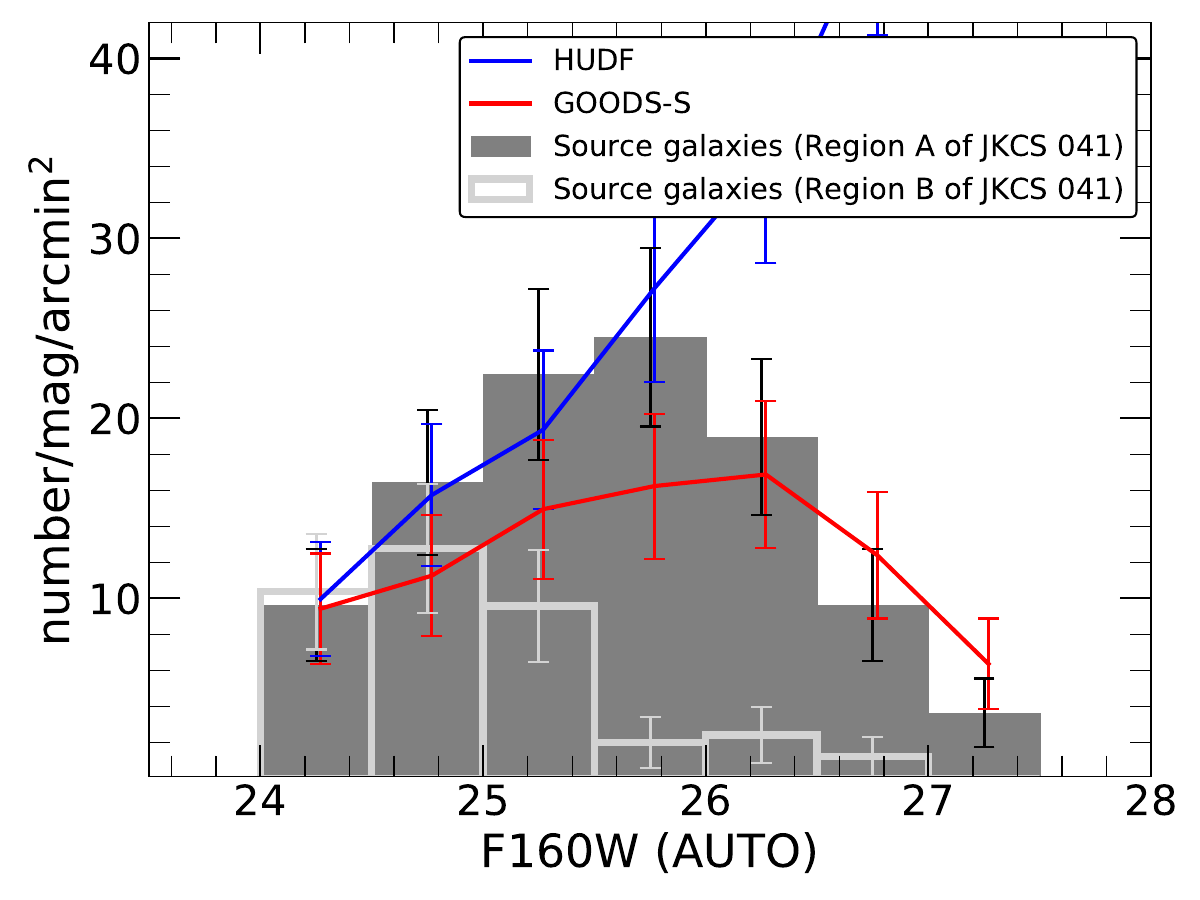}
\includegraphics[width=8.5cm]{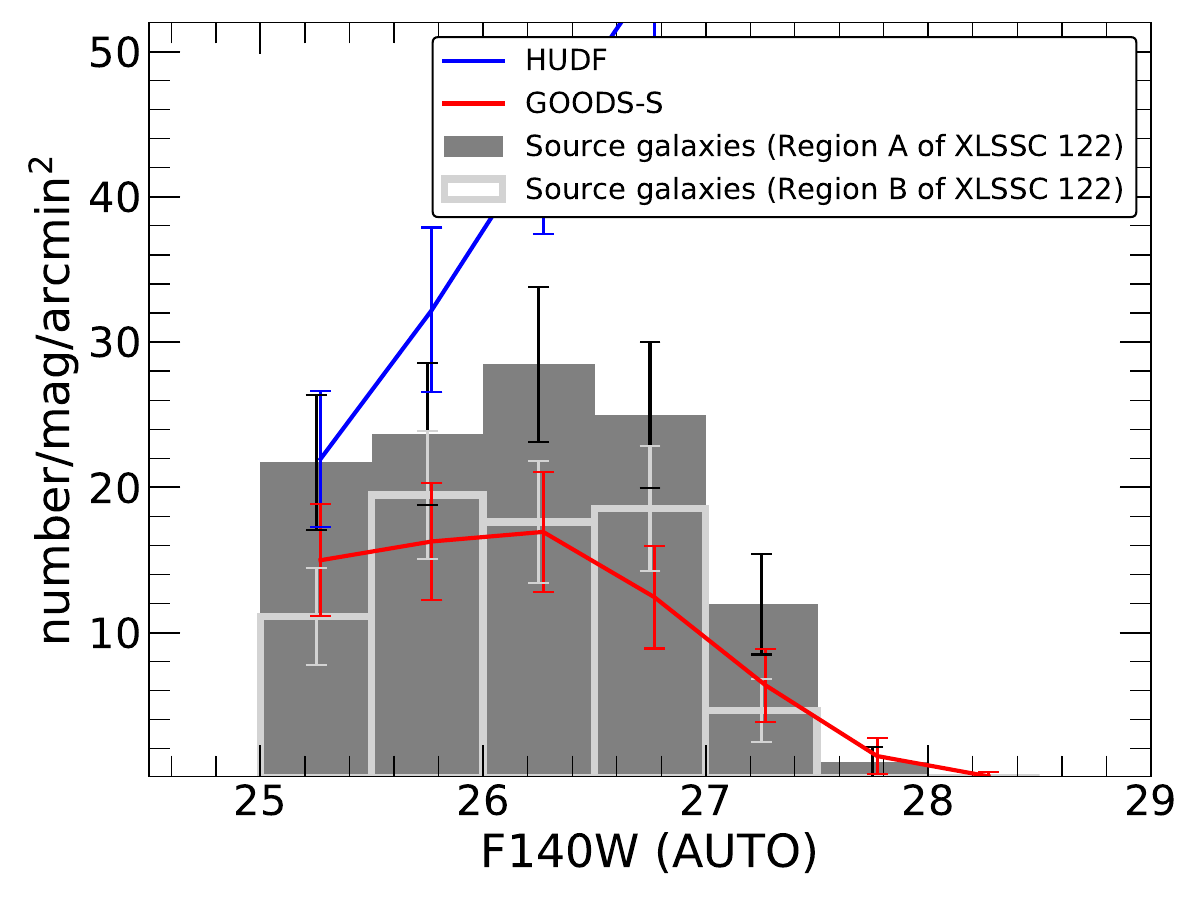}
\caption{Color-magnitude relation and magnitude comparison with control fields for \JKCS~and \XLSS. 
\textbf{Top:} Color-magnitude relation for \JKCS~(left) and \XLSS~(right). The cluster members are indicated in red. Both clusters exhibit a well-defined locus of the red sequence. 
\textbf{Bottom:} Comparison of magnitude distributions between the source population in the cluster fields and those in the control fields.
Error bars are computed based on Poissonian statistics. Given the good consistency between our source galaxies and those in the control fields in region A, the contamination from possible members is insignificant. The rapid drop in the source magnitude distribution in region B is due to the shallow depth.}
\label{fig:cmd}
\end{figure*}

\subsection{Weak-Lensing Analysis} \label{section_WL}
\subsubsection{Modeling Point Spread Function and Measuring Galaxy Shapes} \label{shape_measurement}
The first step in our WL analysis is modeling the PSFs of source galaxies. Incorrect PSF modeling introduces additional distortion, thereby mimicking the WL signal. This systematic effect, resulting from PSF modeling errors, is particularly problematic for faint and small galaxies, comprising the majority of the source population in high-redshift galaxy cluster lensing. Following the methodology outlined in \cite{Jee2007}, we generated PSF libraries for the WFC3/F140W and WFC3/F160W filters based on archival globular cluster field imaging data. These PSF libraries are then employed to identify the best-matching PSF template for each science exposure in the \JKCS~and \XLSS~fields. For detailed descriptions of the \HST~PSF modeling, readers are referred to \cite{Jee2007, Jee2017}. We were unable to create a PSF library for the WFC3/F105W filter due to the limited number of suitable stellar fields in the archive.

We conducted the shape measurement of each object by fitting a two-dimensional (2D) elliptical Gaussian profile using the {\tt MPFIT} \citep{MPFIT} package. Before the fitting, the 2D elliptical Gaussian model is convolved with the PSF model at each source position. The shape parameters of the model (e.g., orientation angle $\phi$, semi-major $a$, and -minor $b$ axes) are utilized to determine the ellipticity $e = (a-b)/(a+b)$. We refer readers to \cite{Kim2019, Kim2021} for a detailed procedure of shape measurement.

In principle, the average ellipticity of the sources corresponds to the reduced shear $g$ defined by $g = \gamma / (1 - \kappa)$, where $\kappa$ is convergence, and $\gamma$ is shear (see a review by \citealt{Bartelmann2001}). However, the reduced shear does not exactly equal the averaged ellipticity due to systematic effects, including model bias and noise bias (e.g., \citealt{Refregier2012}; \citealt{Massey2013}). Here we adopted a multiplicative correction factor of 1.25 determined by \cite{Jee2017}. This value combines the shear multiplicative correction factor for the HST Advanced Camera for Surveys with an additional correction for undersampling bias in WFC3/IR. For further details, we refer the reader to \cite{Jee2017} and \cite{Kim2019}.

\subsubsection{Source Selection and Redshift Estimation} \label{source_selection}
We selected the background source population that satisfies both the shape and photometric conditions and estimated their redshift distributions utilizing control fields, following the description in \cite{Kim2019, Kim2021}. Here, we briefly explain the procedure.

The source selection starts with identifying objects with stable fitting status ({\tt MPFIT STATUS}$ = 1$; \citealt{MPFIT}) in shape measurement. We then removed too small and faint objects, with ellipticity errors larger than 0.25 and semi-minor axes smaller than 0.4 pixels. The remaining point-like sources are further excluded by discarding sources whose half-light radii are smaller than those of stars. Finally, residual spurious objects (e.g., diffraction spikes around bright stars, fragmented parts of foreground galaxies, clipped objects at the field boundaries, etc.) were manually identified and discarded.

After applying the shape requirements, we imposed photometric conditions in region A, where color information is available: $24 < \mbox{F160W} < 28$ with $\mbox{F105W}-\mbox{F160W}< 1.4$ and $25 < \mbox{F140W} < 28$ with $\mbox{F105W}-\mbox{F140W}< 1.6$ for \JKCS~and \XLSS, respectively. These criteria enable us to remove the cluster bright red-sequence galaxies (top panels of Figure~\ref{fig:cmd}). We also excluded the spectroscopic members using the publicly available catalogs from the literature (e.g., \citealt{Newman2014}; \citealt{Prichard2017}; \citealt{Willis2020}). For objects without color information (in region B), only magnitude conditions: $24 < \mbox{F160W} < 28$ and $25 < \mbox{F140W} < 28$ for \JKCS~and \XLSS, respectively, were applied.

We examined possible residual contamination from blue cluster members and faint red sequence galaxies in our source selection because we only removed spectroscopically confirmed blue members. We compared our source magnitude distributions with those of the two control fields: the Hubble Ultra Deep Field (HUDF; \citealt{UVUDF}) and the Great Observatories Origins Deep Survey (GOODS; \citealt{GOODSmag}) after applying the same photometric conditions. The bottom panels of Figure~\ref{fig:cmd} show that our source magnitude distribution in both clusters does not show any significant excess with respect to the ones from the HUDF at $\mbox{F160W} < 26$ or $\mbox{F140W} < 25.5$, which implies that the member contamination is negligible. The rapid drop of source number density at $\mbox{F160W} > 26$ and $\mbox{F140W} > 27$ stems from the fact that our observations are shallower than the HUDF; the $5\sigma$ limiting magnitudes are $27.4$ and $28.8$ for the region A in \JKCS~and the HUDF, respectively. The source number densities in region B are smaller than those in region A and drop quickly because the region is shallower; the $5\sigma$ limiting magnitude is $26.4$ in \JKCS.

From the source selection described above, 766 galaxies, corresponding to a number density of $\mytilde87$~\persqarcmin, are selected for \JKCS. Due to the difference in the observation depth, the source density is higher in region A with $\mytilde106$~\persqarcmin~and $\mytilde38$~\persqarcmin~ for regions A and B, respectively. In the \XLSS~field, 592 galaxies ($\mytilde104$~\persqarcmin) are selected with $\mytilde112$~\persqarcmin~and $\mytilde71$~\persqarcmin~for regions A and B, respectively.

We estimated the lensing efficiency $\beta$ of our source population by employing the HUDF photo-$z$ catalog after applying the same source selection conditions. Following the procedure in \cite{Kim2019, Kim2021}, we find that the average lensing efficiencies $\left < \beta \right >$ of source galaxies are $\left < \beta \right > = 0.079$ (corresponding to $z_{\rm eff}=2.085$) and $\left < \beta \right >=0.100$ (corresponding to $z_{\rm eff}=2.180$) for regions A and B, respectively, in the \JKCS~field. Moreover, we accounted for the width of the source distribution since the assumption that all sources are at the same redshift plane causes systematic errors. The width of the source population is estimated by squaring the lensing efficiency of each source and taking the average, yielding $\left < \beta^{2} \right > = 0.019$ and $0.023$ for regions A and B, respectively. We used the lensing efficiency $\left < \beta \right >$ and its width $\left < \beta^{2} \right >$ to apply the first-order correction derived by \cite{Seitz1997}. For the source population in \XLSS, the lensing efficiency, its width, and effective source redshift in region A (region B) are $\left < \beta \right >=0.063$ (0.094), $\left < \beta^{2} \right >=0.014$ (0.021), and $z_{\rm eff}=2.234$ (2.384), respectively.

\section{Results} \label{section_results}
\subsection{Mass Reconstruction} \label{mass_map}
\begin{figure}
\centering
\includegraphics[width=85mm]{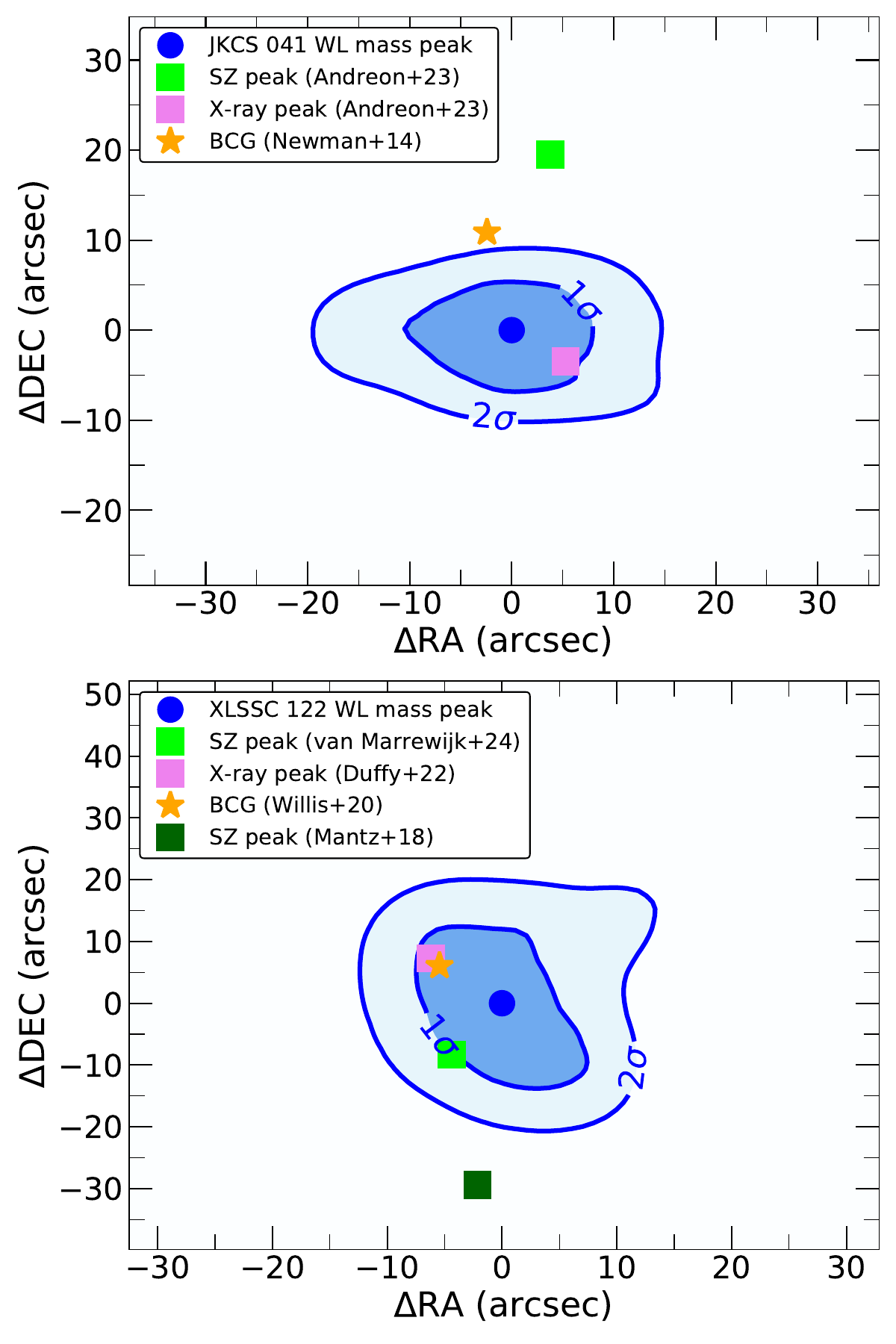}
\caption{
WL mass centroid significance distributions for \JKCS~(top) and \XLSS~(bottom), based on 1000 bootstrap realizations. The centroid distributions for each cluster are represented by blue contours. 
For \JKCS, the X-ray peak is enclosed within the $1\sigma$ region, whereas the BCG lies slightly beyond the $2\sigma$ contour. The SZ peak is offset from the WL mass centroid at the $\mytilde3\sigma$ level. 
For \XLSS, the BCG, X-ray peak, and the SZ peak reported by \cite{vanMarrewijk2024} are all consistent with the WL mass centroid within the $1\sigma$ level. In contrast, the SZ peak reported by \citet{Mantz2018} exhibits an offset of $\mytilde2.8\sigma$.
}
\label{fig:centroids}
\end{figure}

We reconstruct the 2D projected mass distribution with the {\tt FIATMAP} code, which implements the KS93 \citep{KS1993} Fourier inversion in real space \citep{FIATMAP}. 
We refer readers to \cite{Wittman2023} for a brief explanation of the code. We verify that the result is highly consistent with the KS93 inversion result. To determine the noise level, significance, and centroid uncertainty in our mass map, we employ the bootstrapping resampling method. We first resample our source galaxies 1000 times, allowing for redundancy. For each resampling, the corresponding mass map is generated and saved. We then create an uncertainty (rms) map from the resulting 1000 mass maps. A signal-to-noise ratio map is obtained by dividing the mean mass map by the rms map. The mass peak centroid uncertainty is determined from the distributions of the mass peak positions measured from the 1000 realizations (Figure~\ref{fig:centroids}).

We display our mass reconstruction results for two clusters in Figures~\ref{fig:mass_map_JKCS041} and~\ref{fig:mass_map_XLSSC122}. In \JKCS, the mass centroid is $\mytilde7\arcsec$, $\mytilde11\arcsec$, and $\mytilde20\arcsec$ away from the X-ray peak, BCG, and SZ peak, respectively. 
Based on the WL mass centroid significance contours in the top panel of Figure~\ref{fig:centroids}, the X-ray peak falls within the $1\sigma$ contour, while the BCG lies slightly outside the $2\sigma$ contour. 
The distance between the SZ peak and our WL mass peak corresponds to $\mytilde3\sigma$ of our WL mass centroid uncertainty. Recently, \citet{Andreon2023} argued that the offset between the X-ray and SZ peaks is one of the indicators that \JKCS~is undergoing a major merger event. 
In \XLSS, the mass centroid is $\mytilde15\arcsec$, $\mytilde13\arcsec$, and $\mytilde7\arcsec$ offset from the X-ray peak, BCG, and SZ peak, respectively. According to the centroid distribution in the bottom panel of Figure~\ref{fig:centroids}, the WL mass peak is spatially consistent with all three reference positions within the $1\sigma$ range\footnote{The first SZ centroid measurement of the cluster by the CARMA data \citep{Mantz2018} is located $\mytilde24$\arcsec~($\mytilde210$ kpc) south of the WL mass centroid, corresponding to $\mytilde2.8\sigma$ distance.}.

The mass peak significances reach $\mytilde3.7\sigma$ and $\mytilde3.2\sigma$ for \JKCS~and \XLSS, respectively. The east-west elongation in the \JKCS~mass map somewhat resembles the distribution of the cluster galaxies and X-ray emission, while the north-south elongation in \XLSS~ is also reminiscent of the distribution of the galaxy and gas components. However, given the noise level of the current WL data, it is difficult to draw definitive conclusions about these observed patterns.

Scrutiny of our mass maps suggests that there are some hints of substructures in both clusters. In \JKCS, we identify a mass clump in the southwestern (SW) region with a peak significance of $\mytilde3.3\sigma$, located $\mytilde39$\arcsec~($\mytilde334$ kpc) from the X-ray peak. This SW clump is close to the southwest galaxy groups mentioned by \cite{Prichard2017}. In addition, \cite{Newman2014} reported the presence of a foreground structure at $z=1.13$ near the SW mass clump (cyan cross in Figure~\ref{fig:mass_map_JKCS041}). Hence, firm conclusions cannot be reached until deeper WL data become available. Nevertheless, we investigate whether the SW clump is associated with a structure at a significantly lower redshift. This test is carried out by changing our source selection criteria in such a way that the source population includes a large number of the \JKCS~members. If the mass substructure comes from a much lower ($z\ll1.8$) redshift, the \JKCS~member galaxies are legitimate background galaxies. When we select brighter source galaxies (e.g., $22 < \mbox{F160W} < 26$) including the \JKCS~red sequence, the SW clump nearly disappears, while the main clump becomes weaker but still visible. Our test suggests that the SW substructure in our mass map may not be associated with the $z=1.13$ structure mentioned by \cite{Newman2014}. 
In \XLSS, the NW mass peak $\mytilde40\arcsec$ away from the main mass peak has a significance of $\mytilde2.6\sigma$. It is difficult to associate this substructure with any galaxy overdensity.

\subsection{Mass Estimation} \label{mass_estimates}
We derive the mass of each cluster using a parametric approach, assuming that the mass profile of each cluster follows a single spherical Navarro-Frenk-White (NFW; \citealt{NFW1997}) profile. The NFW profile has two free parameters: concentration ($c$) and mass ($M$). Since the two parameters are highly degenerate, mass measurements based on X-ray (e.g., \citealt{Amodeo2016}; \citealt{Duffy2022}) and WL analysis (e.g., \citealt{Hoekstra2015}; \citealt{Martinet2016}; \citealt{Jee2017}; \citealt{Zohren2022}) often assume a mass-concentration ($M-c$) relation to enable stable fitting. In this study, we adopt the mass-concentration ($M-c$) relation of \cite{DJ19} (hereafter DJ19). However, it is worth mentioning that because both the NFW profile and the $M-c$ relation describe average properties, the density profile of an individual cluster may significantly deviate from the average. We address the impact of this so-called model bias in \S\ref{mass_systematics}.

Our $\chi^{2}$ function is expressed as follows:
\begin{equation}
\chi^{2} = \sum_{i} \sum_{j=1,2} \frac{ [ g^{\rm model}_j - g^{\rm obs}_j]^2 } {\sigma_{\rm SN}^2 + \sigma_{e,i}^2}, \label{eqn_model}
\end{equation}
\noindent
where
\begin{equation}
g^{\rm model}_j = g^{\rm model}_j(x_i,y_i,z_s,M,c,x_{\rm c},y_{\rm c}), \label{gT_model}
\end{equation}
\noindent
and $g^{\rm obs}_j = g^{\rm obs}_j(x_i,y_i)$ are the $j$th components of the predicted and observed reduced shears of the $i$th source, respectively, at the position $(x_i,y_i)$ at the source redshift $z_s$, given the cluster mass $M$ centered at ($x_c$, $y_c$) with the concentration $c$. The quantities $\sigma_{\rm SN}$ and $\sigma_{e, i}$ are the ellipticity dispersion (shape noise) and measurement uncertainty, respectively. We set $\sigma_{\rm SN}$ to 0.25. The predicted reduced shear $g^{model}$ follows a spherical NFW profile and the DJ19 $M-c$ relation. We fit our model to the shapes of individual sources without binning by minimizing the $\chi^{2}$ function in Equation~\ref{eqn_model}. We choose the X-ray peak of each cluster as the fiducial centroid for mass measurement. The WL signal within $r<15\arcsec$ is discarded to minimize a number of issues, including cluster member contamination, centroid uncertainty, nonlinearity, etc.

The bottom right panels of Figure~\ref{fig:mass_map_JKCS041} and~\ref{fig:mass_map_XLSSC122} show the observed tangential shear profiles and the best-fit NFW model prediction. The resulting masses of \JKCS~and \XLSS~are $M_{200c} = (5.4\pm1.6)~\times$~\solarm~[$M_{500c} = (3.8\pm1.1)~\times$~\solarm] and $M_{200c} = (3.3\pm1.8)~\times$~\solarm~[$M_{500c} = (2.3\pm1.3)~\times$~\solarm], respectively. The $M_{500c}$ value is obtained by interpolating from the $M_{200c}$ value, assuming the NFW profile with the DJ19 $M-c$ relation.

To address mis-centering (e.g., \citealt{Martel2014}), we repeat the mass measurements at three additional centroids: the SZ peak, BCG position, and WL mass centroid. The resulting masses for different center choices are highly consistent with the results obtained using the X-ray peaks. Quantitative comparisons are presented in \S\ref{mass_comparison}.

\section{Discussion} \label{section_discussion}
\subsection{Mass Comparison} \label{mass_comparison}
\begin{figure}
\centering
\includegraphics[width=8.5cm]{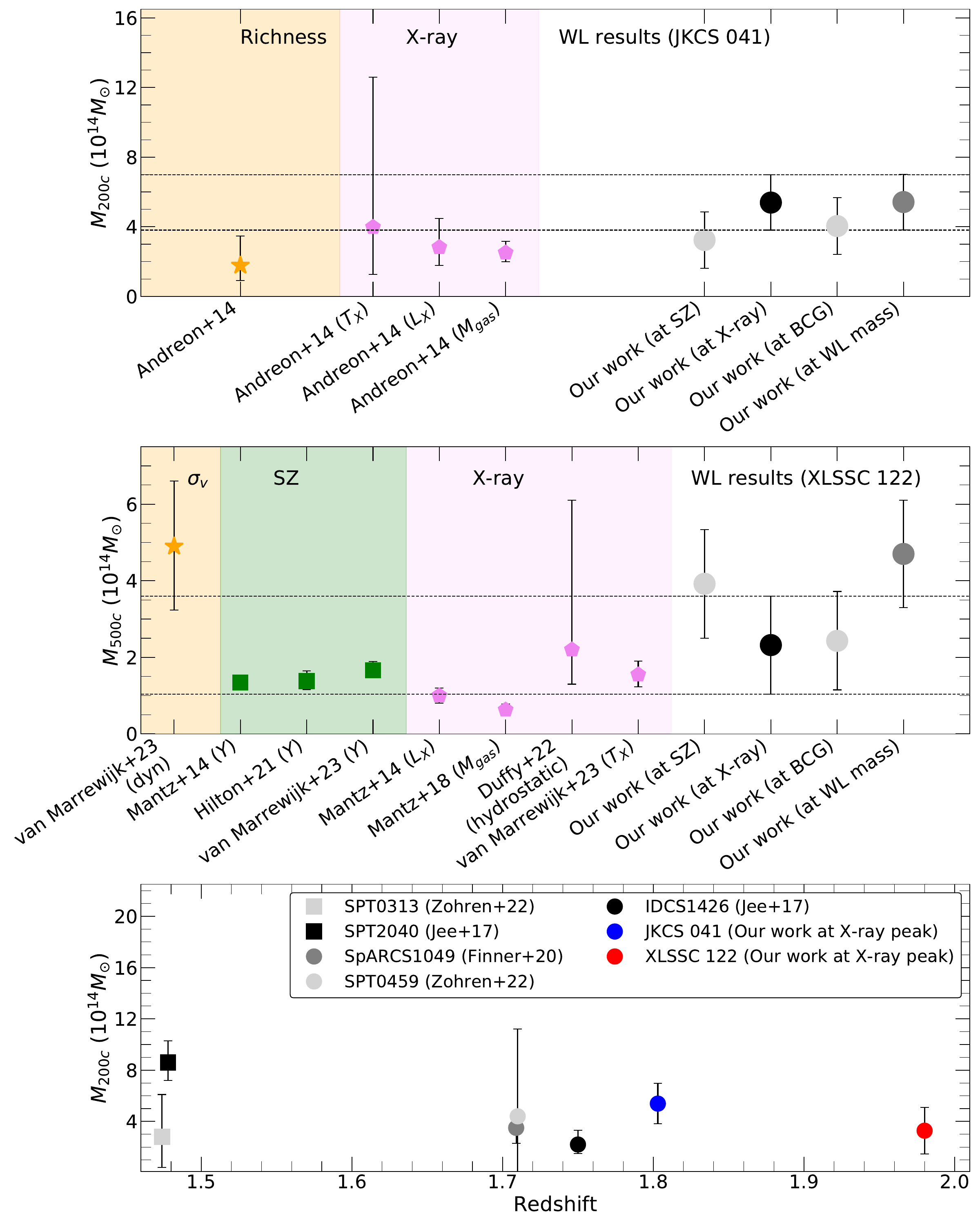}
\caption{Mass comparison of \JKCS~(top) and \XLSS~(middle) with various proxies and other high-$z$ ($z \gtrsim 1.5$) clusters with WL mass measurements reported (bottom). For our WL mass representation, we adopt the best-fit NFW results with \cite{DJ19} $M-c$ relation under a single spherical halo assumption centered at the SZ, X-ray, BCG, and our WL mass centroid. Our WL mass results for both clusters are statistically consistent with previous mass estimates derived from member galaxies and gas components within the 1$\sigma$ uncertainties. Additionally, our sample clusters exhibit similar mass estimates to galaxy clusters at lower redshift (e.g., $z\sim1.7$).}
\label{fig:mass_compare}
\end{figure}

We compare previous mass estimates with our WL masses for \JKCS~and \XLSS~in the top and middle panels of Figure~\ref{fig:mass_compare}, respectively. Note that as only $M_{500c}$ values have been previously reported for \XLSS, we present our $M_{500c}$ values in the middle panel.

The previous mass measurement of \JKCS~was performed by \cite{Andreon2014}, who determined the cluster’s mass using four different mass proxies: richness, X-ray temperature, X-ray luminosity, and gas mass, based on mass-scaling relations. These proxies provided consistent mass estimates with $M_{200} > 10^{14.2}~M_{\sun}$. \cite{Andreon2014} concluded the cluster is massive for its redshift ($z=1.8$). The top panel of Figure~\ref{fig:mass_compare} shows their results with $1\sigma$ uncertainties. Our WL mass estimates, with four different mass centroid choices, are consistent with them.

For \XLSS, several studies have measured the cluster mass using its member galaxies and intracluster medium. The X-ray and SZ mass estimates are within the range $M_{500c} = (1 - 2)\times10^{14}~M_{\sun}$ \citep{Mantz2014,Mantz2018,Hilton2021,vanMarrewijk2024}. Recently, \cite{vanMarrewijk2024} suggested that the mass estimate based on velocity dispersion is $\mytilde2.5$ times greater than those derived from the hot gas. 
Our WL masses, measured with the X-ray and BCG centroids, are $M_{500c}\sim2\times10^{14}~M_{\sun}$, consistent with the gas-based mass estimates, while their $1\sigma$ error bars touch the $1\sigma$ lower limit of the dynamical mass. The WL masses obtained with the SZ and mass centroids are $M_{500c} = (4 - 5)\times10^{14}~M_{\sun}$, closer to the dynamical mass.

In the bottom panel of Figure~\ref{fig:mass_compare}, we compare our fiducial WL masses with five other high-redshift clusters at $z \gtrsim 1.5$: SPT-CL J0313-5334 ($z = 1.47$, SPT0313) and SPT0459 ($z = 1.71$) from \cite{Zohren2022}, SPT-CL J2040-4451 ($z = 1.48$, SPT2040) and IDCS1426 ($z = 1.75$) from \cite{Jee2017}, and SpARCS1049 ($z = 1.71$) from \cite{Finner2020}. The most massive cluster at $z \gtrsim 1.5$ to date is SPT2040 at $z=1.48$, whose mass is $M_{200}\sim 8 \times10^{14}~M_{\sun}$ \citep{Jee2017}. Apart from SPT2040, however, our two clusters, \JKCS~and \XLSS, are comparable in mass to the other four clusters. Since the cluster mass function at its massive end decreases rapidly from $z=1.5$ to $z=2$, it is interesting to examine whether or not we can accommodate massive clusters such as \JKCS~and \XLSS~at $z\sim2$ within the current $\Lambda$CDM paradigm. We address this issue in \S\ref{rarity}.

\begin{figure}
\centering
\includegraphics[width=8.5cm]{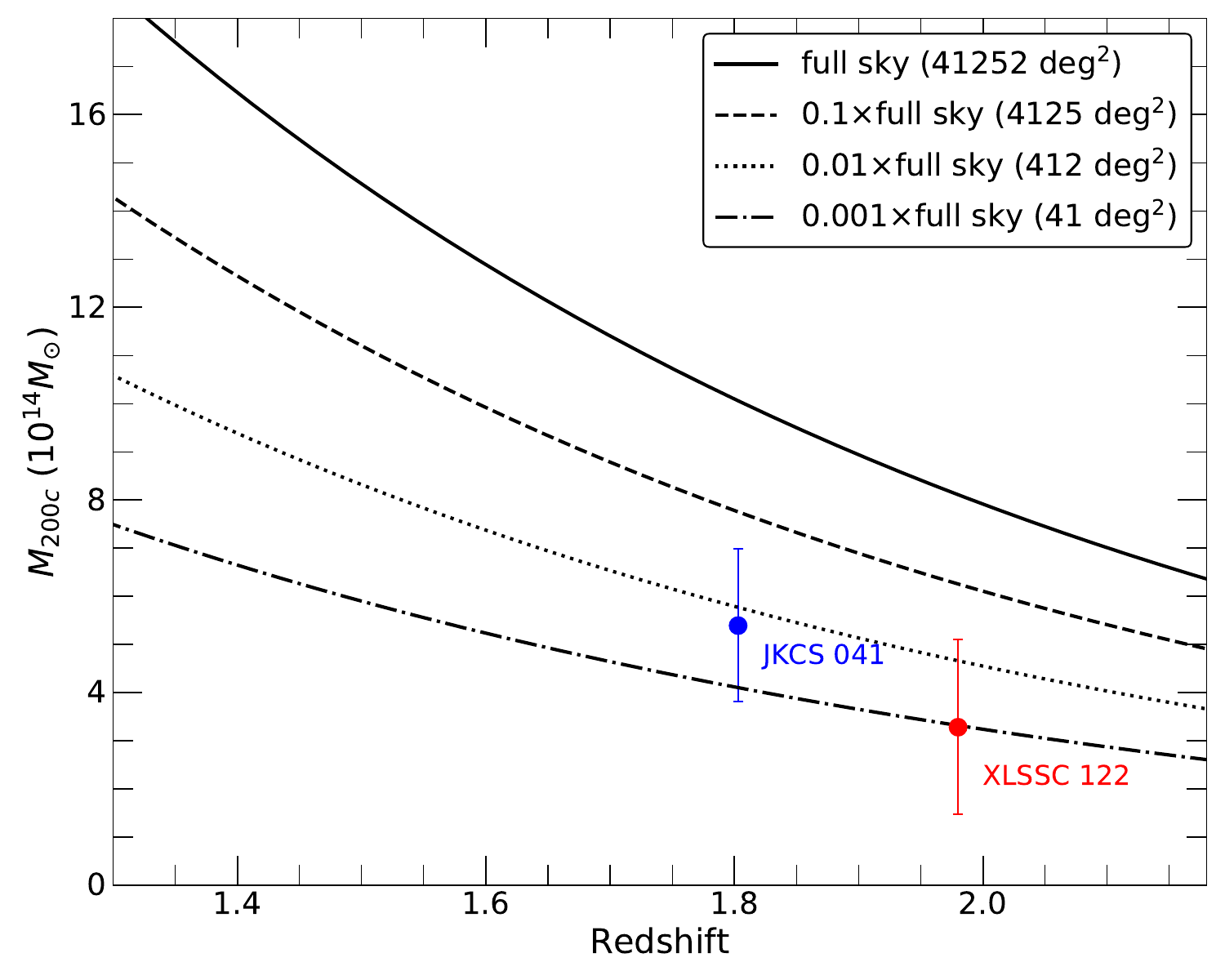}
\caption{Exclusion curve and mass estimates of \JKCS~and \XLSS. The exclusion curve considers sample and parameter variances within a 95\% CL ($\mytilde2\sigma$) for the full sky ($\mytilde41,000$~deg$^2$, solid), $\mytilde4,100$~deg$^2$ (dashed), $\mytilde410$~deg$^2$ (dotted), and $\mytilde41$~deg$^2$ (dashed-dotted). 
}
\label{fig:rarity}
\end{figure}

\begin{figure}
\centering
\includegraphics[width=85mm]{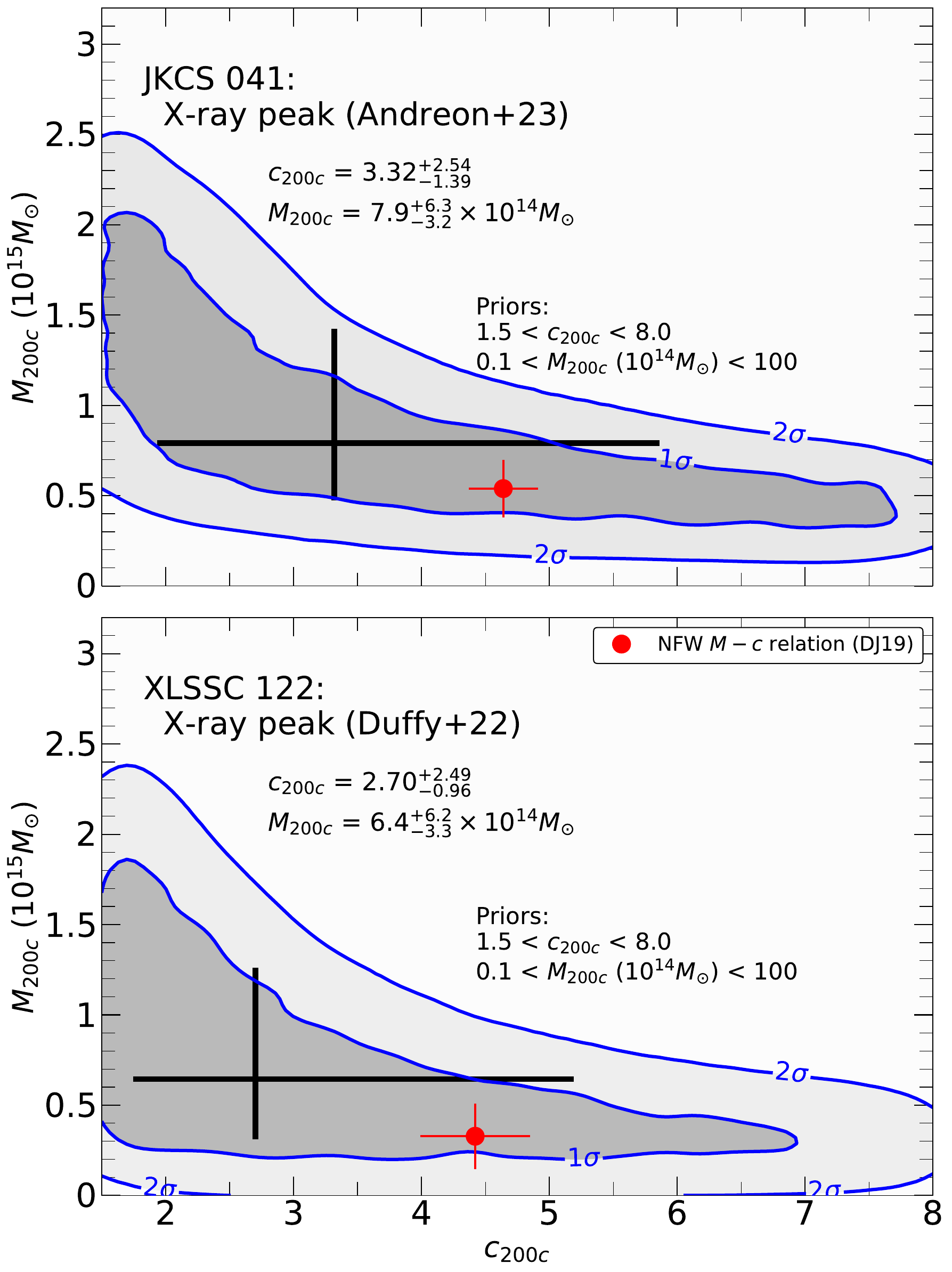}
\caption{
Posterior distributions based on 250,000 MCMC samples centered at the X-ray peaks for \JKCS~(top) and \XLSS~(bottom), derived without assuming an $M-c$ relation. The black crosses indicate the best-fit values and the associated uncertainties of $c_{200c}$ and $M_{200c}$, obtained through one-dimensional marginalization. For comparison, the results obtained using the DJ19 $M-c$ relation are also shown. The mass estimates from the MCMC samples are in good agreement with those derived using the DJ19 relation. }
\label{fig:MCMC_figs}
\end{figure}

\subsection{Rarity} \label{rarity}
We study the rarity of \JKCS~and \XLSS~using the exclusion curve test suggested by \cite{Mortonson2011}. The exclusion curve represents the locus of the predicted maximum mass across redshift for a given cosmology and survey volume. It also accounts for parameter and sample variances of the cosmological model within a specified confidence level (CL). In Figure~\ref{fig:rarity}, we present exclusion curves at the 95\% CL for both sample variance and parameter uncertainties for the areas of $41,000$~deg$^2$ (full sky), $4,100$~deg$^2$, $410$~deg$^2$, and $41$~deg$^2$. A cluster with a mass and redshift above the curve for a given volume is unlikely to be observed within the corresponding survey under the \LCDM~model.

Our WL masses of the two clusters are comfortably below the exclusion curves for the $41,000$~deg$^2$ and $4,100$~deg$^2$ surveys, and overlap with the curves for the $410$~deg$^2$ and $41$~deg$^2$ surveys. It is important to note that our WL mass results shown here have not been corrected for Eddington bias, which arises from the steep mass function \citep{Eddington1913}. Applying the Eddington bias correction (e.g., \citealt{Andreon2009}; \citealt{Mortonson2011}) would reduce the mass, increasing the compatibility of these two systems with \LCDM~cosmology.

\subsection{Mass Systematics} \label{mass_systematics}
In \S\ref{mass_estimates}, we determine the masses of the two high-$z$ clusters assuming the NFW halo profile with the DJ19 $M-c$ relation. While WL is generally considered superior to other methods that require assumptions about hydrostatic equilibrium or scaling relations, it is important to note that WL mass estimation is subject to other systematics.

The $M-c$ relation reflects the averaged properties of simulated galaxy clusters. Also, \citet{Lee2023} demonstrated that the mass and concentration of merging clusters vary depending on their merger phase. Additionally, the spherical NFW profile, an average representation from numerical simulations, does not account for the potentially inhomogeneous shapes of dark matter halos. Several studies (e.g., \citealt{Clowe2004}; \citealt{Meneghetti2010}; \citealt{Becker2011}; \citealt{Gruen2015}) have explored the mass uncertainties arising from deviations from spherical symmetry, such as halo triaxiality and orientation relative to the observer. High-$z$ clusters like \JKCS~and \XLSS, which are actively growing, may deviate from spherical symmetry and exhibit scatter in the $M-c$ relation.

To examine the impact of model bias from the $M-c$ relation and the assumption of a spherical NFW halo, we perform experiments using Markov Chain Monte Carlo (MCMC) analysis and aperture mass densitometry (AMD). The MCMC analysis allows us to probe the $M-c$ parameter space freely without being constrained by a tight relation between the two parameters. If the resulting posteriors are significantly different from the result obtained with the $M-c$ relation, this indicates that the systems may significantly deviate from the assumed relation.We employ flat priors for $c_{200c}$ and $M_{200c}$ with the intervals $1.5 < c_{200c} < 8$ and $10^{13} M_{\sun} < M_{200c} < 10^{16} M_{\sun}$, respectively. 
Figure~\ref{fig:MCMC_figs} present the posterior distributions for the mass and concentration parameters from 250,000 MCMC samples for \JKCS~and \XLSS. The marginalized masses of \JKCS~and \XLSS~are estimated to be $M_{200c} = 7.9^{+6.3}_{-3.2}~\times$~\solarm~and $M_{200c} = 6.4^{+6.2}_{-3.3}~\times$~\solarm, respectively. Although the median values are somewhat higher, the posteriors overlap with the results obtained with the $M-c$ relation.

The AMD approach provides a projected mass profile without the need for a specific halo model. For a detailed description of this method, readers are referred to \citet{Clowe2000} and \citet{Jee2005}. We find that the non-parametric AMD aperture mass aligns well with the parametric NFW aperture mass at $R_{2500c}$; because of the limited field size, the comparison is not feasible significantly beyond $R_{2500c}$, corresponding to $\mytilde34$\arcsec~($\mytilde290$ kpc) and $\mytilde27$\arcsec~($\mytilde230$ kpc) for \JKCS~and \XLSS, respectively. 
The 2D projected masses from the best-fit NFW models within the $r=R_{2500c}$ aperture are $M_{{\rm proj},2500c} = (2.5\pm0.5)~\times$~\solarm~and $M_{{\rm proj},2500c} = (1.5\pm0.6)~\times$~\solarm~for \JKCS~and \XLSS, respectively. These values are consistent with the AMD aperture masses: $M_{{\rm proj},2500c} = (2.6\pm0.5)~\times$~\solarm~and $M_{{\rm proj},2500c} = (2.0\pm0.4)~\times$~\solarm~for \JKCS~and \XLSS, respectively.

Finally, we consider large-scale structures along the line of sight that are uncorrelated with our high-$z$ clusters. Following the \cite{Hoekstra2003} recipe, we estimate that large-scale structures contribute to $\sigma_{\gamma}\sim0.01$ within the angular scale that we probe \citep{Jee2017}, which corresponds to $15 - 20$\% error in cluster mass estimation.

In summary, the above experiments indicate that the total error budgets in cluster mass are dominated by the statistical uncertainties for both clusters. However, given that the HST field is not sufficiently large enough to cover the virial radii of the clusters, it is necessary to revisit the issue with wider imaging data in a future study.

\begin{figure}
\centering
\includegraphics[width=8.5cm]{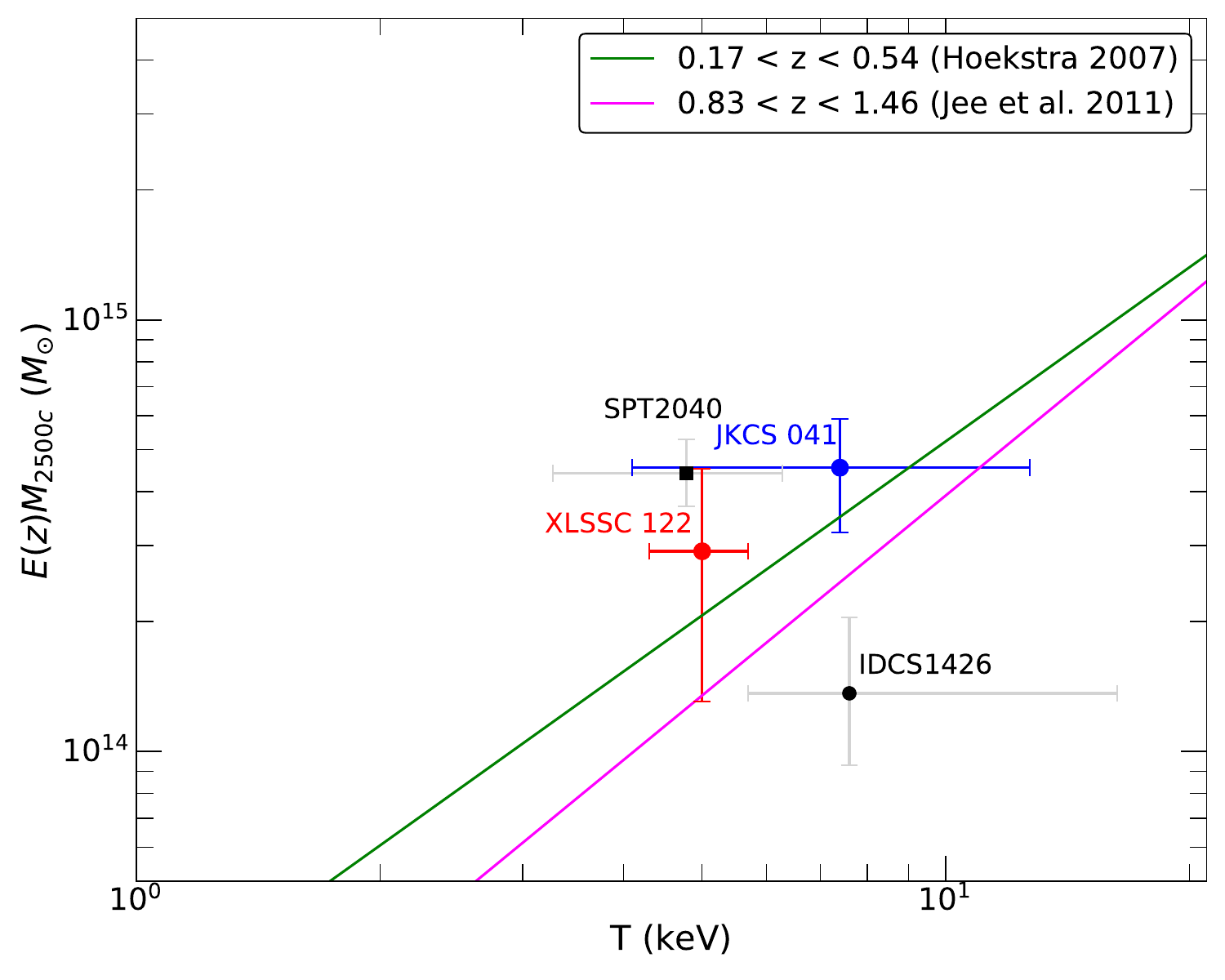}
\caption{WL mass vs. X-ray temperature relation of our sample (\JKCS~and \XLSS) and other high-$z$ ($z \gtrsim 1.5$) clusters with X-ray and WL mass reported in the literature: WL masses from \cite{Jee2017} and X-ray temperatures from \cite{Bulbul2019} and \cite{Brodwin2016} for SPT2040 and IDCS1426, respectively. We display two best-fit results of the scaling relations at low-$z$ ($M \propto T^{1.34}$; \citealt{Hoekstra2007}) and high-$z$ ($M \propto T^{1.54}$; \citealt{Jee2011}). 
\JKCS~and \XLSS~follow the scaling relations measured from both low-$z$ and high-$z$ clusters.}
\label{fig:scaling_rel}
\end{figure}

\subsection{X-ray - WL Mass Scaling Relation} \label{scaling_rel}
WL mass serves as a calibrator for mass estimates derived from other mass proxies, such as X-ray and SZ observables. The relationship between mass and observables from gas components in galaxy clusters is typically characterized by the self-similar model \citep{Kaiser1986}. While many studies have established WL mass-observable relations for low-redshift clusters (e.g., \citealt[][]{Hoekstra2007, Hoekstra2015}; \citealt{Mulroy2019}), far fewer have explored these relations at high redshifts (i.e., $0.8 < z < 1.7$; \citealt{Jee2011}; \citealt[][]{Schrabback2018, Schrabback2021}; \citealt{Zohren2022}). Given that the two clusters in this study represent the most distant galaxy cluster sample where WL signals have been detected, we examine the locations of \JKCS~and~\XLSS~in the WL mass and X-ray temperature plane.

Figure~\ref{fig:scaling_rel} shows the locations of \JKCS~and XLSSC 122 in the WL mass and X-ray temperature plane. We also display the relations for low- \citep{Hoekstra2007} and high-redshift \citep{Jee2011} clusters from previous studies. For a consistent comparison with previous results, we convert our $M_{200c}$ masses to $M_{2500c}$, assuming the NFW profile with the DJ19 $M-c$ relation. We utilize the X-ray temperature measurements from \cite{Andreon2014} for \JKCS~and from \cite{Mantz2018} for \XLSS. Together with \JKCS~and~\XLSS, we also include two other clusters at $z \gtrsim 1.5$, whose X-ray and WL mass measurements are available in the literature. \JKCS~and~\XLSS~follow the scaling relations of \cite{Hoekstra2007} and \cite{Jee2011} well, while the other two clusters are marginally consistent with these scaling relations.

\section{Summary} \label{section_summary}
We present a WL study of the two high-redshift clusters \JKCS~and \XLSS~at $z=1.80$ and $1.98$, respectively. They are the two most distant galaxy clusters to date ever measured with WL. Using the \HST~WFC3/IR imaging data and careful WL analysis, we successfully detected the WL mass peaks of \JKCS~and \XLSS~at the $\mytilde3.7\sigma$ and $\mytilde3.2\sigma$ levels, respectively. 
The mass peak positions show good agreement with the X-ray peaks of the respective clusters.

Assuming a single spherical NFW halo following the DJ19 $M-c$ relation centered on the X-ray peak, we determine the masses of \JKCS~and \XLSS~to be $M_{200c} = (5.4\pm1.6)~\times$~\solarm~and $M_{200c} = (3.3\pm1.8)~\times$~\solarm, respectively. Our MCMC and AMD experiments suggest that the model bias due to the spherical NFW with the $M-c$ relation may be subdominant with respect to the statistical uncertainty for both clusters.

Our exclusion curve test shows that although the two clusters are certainly extremely massive for their redshifts, their existence can comfortably be accommodated within the current \LCDM~paradigm. Finally, we conclude that \JKCS~and \XLSS~well follow the mass-temperature scaling relations in the literature. \\

M.J.J. acknowledges support for the current research from the National Research Foundation (NRF) of Korea under the programs 2022R1A2C1003130 and RS-2023-00219959. J.K. and L.M. acknowledge support from the Beecroft Trust. We thank the anonymous referee for suggestions that improved the manuscript. This work is also based on observations made with the NASA/ESA Hubble Space Telescope (HST). The data presented in this paper were obtained from the Mikulski Archive for Space Telescopes (MAST) at the Space Telescope Science Institute. The specific observations analyzed can be accessed via \dataset[doi:10.17909/tksy-qb78]{https://doi.org/10.17909/tksy-qb78}.

\software{ 
MultiDrizzle \citep{2002multidrizzle}, 
DrizzlePac \citep{DrizzlePac}, 
SExtractor \citep{Bertin1996}, 
{\tt MPFIT} \citep{MPFIT}, 
{\tt FIATMAP} \citep{FIATMAP}, 
}

\bibliographystyle{aasjournal}

\end{document}